\documentclass[aps,prfluids]{revtex4-2}
\usepackage{amssymb}
\usepackage{graphicx}
\usepackage{bm}
\usepackage{physics}

\usepackage{hyperref}
\hypersetup{
    colorlinks = true,
    urlcolor   = blue,
    citecolor  = black,
}

\newcommand{\dInt}{\mathrm{d}}

\begin{document}

\title{Asymptotic limits of the axisymmetric solution of the Brinkman equation for a point force near a no-slip wall} 

\author{Abdallah Daddi-Moussa-Ider}
\email{admi2@open.ac.uk}
\affiliation{School of Mathematics and Statistics, The Open University, Walton Hall, Milton Keynes MK7 6AA, United Kingdom}
\author{Andrej Vilfan}
\email{andrej.vilfan@ijs.si}
\affiliation{Jo\v{z}ef Stefan Institute, 1000 Ljubljana, Slovenia}
\date{\today}

\begin{abstract}
We derive the far-field and near-field solutions for the Green’s function of a point force acting perpendicular to a no-slip wall in a Brinkman fluid, focusing on the regime where the distance between the force and the wall is much smaller than the screening length. The general solution is obtained in closed form up to a single integral and can be systematically expanded in a Taylor series in both the far-field and near-field limits. The flow can then be expressed as a series of source-multipole singularities with an additional, analytically known, correction in the proximity of the wall. Comparisons with numerical integration demonstrate the accuracy and reliability of the asymptotic expansions. The results are also applicable to the unsteady Stokes flow driven by a localized assembly of forces, such as a beating cilium protruding from a flat surface. 
\end{abstract}

\maketitle

\section{Introduction}
\label{sec:intro}

The Brinkman equation, also known as Brinkman-Debye-Bueche (BDB) equation~\cite{brinkman1949permeability, brinkman1949permeability2, Debye.Bueche1948}, describes the effective flow of a fluid through a porous medium~\cite{ingham1998transport}.
The Brinkman description represents an extension of the classical form of Darcy's law~\cite{auriault2009domain} and was originally formulated to model the viscous flow past a sparse array of identical non-overlapping spherical particles~\cite{happel1958viscous, childress1972viscous, howells1974drag, durlofsky1987analysis}. It becomes important when the length scale of the flow gradients becomes comparable to the screening length in the porous medium. Brinkman equation can also be viewed as the Stokes equation with an additional local drag force density on the permeating fluid that is negatively linear in the fluid velocity. The Brinkman description is also naturally linked to unsteady Stokes flow and linear viscoelastic fluids by a correspondence principle~\cite{kim13}. 

In a Stokes fluid, the Green's function in the presence of a wall with a no-slip boundary belongs to the oldest and most fundamental solutions. It was first formulated by Lorentz~\cite{lorentz07} and later represented by Blake as a multipole image~\cite{blake1971note}, which gave it the now common designation as Blake's tensor. The solution is central in problems where the flow is induced by an object embedded in a planar surface, such as a beating cilium~\cite{Blake1972,Vilfan2012}, or moving in the proximity of such a surface, like a swimming microorganism or an artificial microswimmer~\cite{spagnolie12,ibrahim2016walls,daddi18jcp}. Furthermore, the Green's function provides the basis to determine the leading-order correction to the viscous drag coefficient of a small particle in the proximity of a wall~\cite{happel12}. 

In the case of a Brinkman fluid or unsteady Stokes flow, a general analytical expression for the Green's function does not exist. Pozrikidis has determined the solution in Fourier space, but concluded that the integrals could not be solved in a closed form, which implies that the image system may not be represented with a finite number of wall-image singularities~\cite{pozrikidis1989singularity}. 
Felderhof has derived the solution in a different form in frequency~\cite{felderhof05} and time domain~\cite{felderhof2009flow} and used it to evaluate the reaction fields that determine the wall effect on the mobility of a small particle. For the latter, a closed-form solution was found~\cite{felderhof05}.   
The result has been refined beyond the point-particle approximation~\cite{simha2018unsteady}. 
A review of the existing results is provided, with analytical asymptotic solutions derived for the self-mobility of a spherical particle in the presence of a wall in the limit of distances that are short or long compared to the viscous penetration depth~\cite{fouxon2018fundamental}.
The frequency-dependent mobility of a particle can be directly compared to microhydrodynamic experiments~\cite{jeney08,franosch09}. In addition to point particles, an exact representation of Brinkman flow due to a regularized Brinkmanlet near a plane wall has also been provided in Fourier space~\cite{nguyen2019computation}, but as in the case of a point force Brinkmanlet it requires numerical Fourier transformation to return to real space. Another extension includes replacing the infinite no-slip wall with a finite-sized disk, which has been solved for the axisymmetric case~\cite{DaddiMoussaIder.Golestanian2023}. 
For a force parallel to the wall, the far-field limit and the integral force balance has been discussed~\cite{daddi2025force}.

In this paper, we revisit the problem for the case when the point force is acting perpendicular to the no-slip boundary and is localized close to the wall, at a distance that is much smaller than the screening length. The general solution has a closed form up to one integral, which can be expanded in a Taylor series both in the far-field and the near-field regime. 
The main contribution of this work is therefore the derivation of closed-form analytical far-field and near-field solutions for the flow induced by a point force near a no-slip wall in a Brinkman medium. Earlier studies have discussed this problem but relied on numerical evaluations or integral representations, without obtaining an analytical asymptotic form \cite{FENG_GANATOS_WEINBAUM_1998,CLARKE_COX_WILLIAMS_JENSEN_2005}.

The asymptotic solutions derived here provide clear physical insight into the nature of hydrodynamic singularities in Brinkman flow and reveal how the presence of the no-slip wall and the medium permeability modify the long-range flow structure. Beyond its theoretical significance, this result has practical relevance: it can be applied to any problem where a non-steady force is acting on a fluid in the vicinity of a wall, for example by a beating cilium. Besides active particles, the Green's functions determine the time-dependent cross-correlation functions of Brownian particles near a wall. Further, they can be used as a fundamental building block for studying hydrodynamic interactions in systems where a force multipole, such as a microswimmer, acts on a fluid in the proximity of a wall.

\section{Problem statement}

The flow is governed by the Brinkman equation and the incompressibility condition \cite{brinkman1949permeability}
\begin{equation}
	\eta \boldsymbol{\nabla}^2 \bm{v} - \eta \alpha^2  \bm{v}   -\boldsymbol{\nabla}p + \bm{f}  = \bm{0} \, , \quad
	\boldsymbol{\nabla} \cdot \bm{v} = 0 \, , \label{Brinkman-Eqs}
\end{equation}
where $\eta$ is the dynamic viscosity of the Newtonian fluid, $\alpha^2$~is the impermeability of the porous medium, which has the dimension of (length)$^{-2}$, and $\bm{v}$ and~$p$ are the fluid velocity and pressure, respectively. 
Here, $\bm{f} = \bm{F} \delta \left( \bm{r} - h \hat{\bm{e}}_z \right)$ is a point-force density acting on the surrounding fluid above the no-slip wall at position $(0,0,h)$.
We adopt the system of cylindrical coordinates $(\rho, \theta, z)$, where $\rho$, $\theta$ and~$z$ denote the radial, azimuthal and axial coordinates, respectively.

\section{Hydrodynamic fields}

We express the solution for the hydrodynamic fields in the form
\begin{equation}
	\bm{v} = \frac{F}{8\pi\eta} \, \bm{G} \, , \qquad
	p = \frac{F}{8\pi} \, P \, ,  
\end{equation}
where $\bm{G} = \bm{G}^\infty + \bm{G}^\text{B}$ and $P = P^\infty + P^\text{B}$.
Here, $\bm{G}^\infty$ and~$P^\infty$ denote the free-space contributions to the velocity and pressure fields, respectively, while $\bm{G}^\text{B}$ and~$P^\text{B}$ represent the corresponding correction terms required to satisfy the prescribed no-slip boundary conditions at the wall.

In the absence of forces, a general solution of the Brinkman equation can be decomposed into a pressure driven flow and a solution of the Helmholtz equation:
\begin{equation}
	\bm{G}=-\alpha^{-2} \, \boldsymbol{\nabla} P + \alpha^{-2} \, \boldsymbol{\nabla}\boldsymbol{\nabla} \cdot \boldsymbol{\Psi} -\boldsymbol{\Psi}
\end{equation}
where $\boldsymbol{\nabla}^2 P=0$ and $(\boldsymbol{\nabla}^2 - \alpha^2) \boldsymbol{\Psi}=\bm{0}$. For the free-space Green's function, the two functions read \cite{kim13}:
\begin{equation}
	\label{eq:p-psi-inf}
	P^\infty= \frac{2}{s^3} (z-h) \,  , \qquad \boldsymbol\Psi^\infty=\hat {\bm{e}}_z \Psi^\infty= -\hat {\bm{e}}_z \, \frac{2}{s} \, e^{-\alpha s} \, , 
\end{equation}
with $s=\sqrt{\rho^2+(z-h)^2}$, denoting the distance from the singularity.
The first contribution represents a long-range pressure-driven flow, equivalent to a source-dipole. The second is an exponentially decaying shear-driven flow. 
In this way, the free-space Brinkmanlet is given by~\cite{howells1974drag}
\begin{subequations} \label{eq:free-space-brinkmanlet}
	\begin{align}
		G_\rho^\infty &= B_2 \,  \frac{\rho \left(z-h\right)}{s^3} \, , \\
		G_z^\infty &= \frac{B_1}{s} + \frac{B_2}{s^3} \left(z-h\right)^2 \, , 
	\end{align} 
\end{subequations}
where the two functions of distance $B_1$ and~$B_2$ are defined as
\begin{subequations}
	\begin{align}
		B_1 &= 2e^{-\alpha s} \left( 1 + \frac{1}{\alpha s} + \frac{1}{(\alpha s)^2} \right) - \frac{2}{(\alpha s)^2} \, , \\
		B_2 &= \frac{6}{(\alpha s)^2} - 2 e^{-\alpha s} \left( 1 + \frac{3}{\alpha s} + \frac{3}{(\alpha s)^2} \right) .
	\end{align}
\end{subequations}

After a zeroth-order Hankel transform, the solutions can be expressed in integral form as
\begin{subequations}\label{eq:p-psi-infty-fourier}
	\begin{eqnarray}
		P^\infty &=&  \int_0^\infty
		2k \, e^{-k|z-h|} J_0(k\rho) \, \dInt k \label{eq:PInfty} \\
		\Psi^\infty &=&  - \int_0^\infty
		\frac{2k}{K} \, e^{-K|z-h|} J_0(k\rho) \, \dInt k \, ,
		\label{eq:PsiInfty}
	\end{eqnarray}
\end{subequations}
where we have defined the abbreviation $K = \sqrt{k^2+\alpha^2}$, with $k$ denoting the wavenumber.
Here, $J_0$ represents the zeroth-order Bessel function of the first kind~\cite{abramowitz72}.
Note that the integral defined in Eq.~\eqref{eq:PInfty} can be found in Gradshteyn and Ryzhik~\cite{gradshteyn2014table}, Eq.~6.621.4, whereas Eq.~\eqref{eq:PsiInfty} follows from Eq.~6.646.1 after applying the substitution $x = \sqrt{t^2 + 1}$.

We now use the same decomposition for the image fields. The no-slip boundary condition at the wall requires
\begin{equation}
	\label{eq:bcz}
	G_z=-\alpha ^{-2} \, \partial_z P +\alpha^{-2} \, \partial^2_z \Psi -\Psi=0
\end{equation}
and
\begin{equation}
	\label{eq:bcr}
	G_\rho=-\alpha^{-2} \, \partial_\rho P +\alpha^{-2} \, \partial_\rho \partial_z \Psi=0\;.
\end{equation}
Because both fields vanish at infinity, the radial condition is equivalent to $\left. (P-\partial_z \Psi)\right|_{z=0}=0$. 
The boundary conditions are satisfied with the following image solution 
\begin{subequations}\label{eq:image-p-psi}
	\begin{eqnarray}
		P^\text{B} &=&  \int_0^\infty
		2k \left( -(\kappa+1) e^{-kh} + \kappa e^{-Kh}\right)  e^{-kz} J_0(k\rho) \, \dInt k \label{eq:image-p}\\
		\Psi^\text{B} &=&  \int_0^\infty
		2 \left( -(\kappa+1)\frac k K \, e^{-Kh} + \kappa e^{-kh}\right) e^{-Kz}    J_0(k\rho) \, \dInt k \, , \label{eq:image-psi}
	\end{eqnarray}
\end{subequations}
where we have introduced the dimensionless function 
\begin{equation}
	\kappa = \frac{ 2k \left( k+K\right)}{\alpha^2} \, .
\end{equation}
Note the mixed nature of the image: A source dipole (contribution of $P^\infty$) induces both a pressure-driven (first term in Eq.~\eqref{eq:image-p}) and a shear-driven image (second term in Eq.~\eqref{eq:image-psi}). Conversely, the shear-driven part of the free-space solution ($\Psi^\infty$) induces a pressure-driven (second term in Eq.~\eqref{eq:image-p}) and a shear-driven image (first term in Eq.~\eqref{eq:image-psi}).

\subsection{Limit of small $h$}

The following analysis applies in the limit where the force is close to the boundary such that $\alpha h \ll 1$ and $h \ll r$, where $r=\sqrt{\rho^2+z^2}$.  As in the case of the Stokes fluid and a force perpendicular to the wall \cite{blake1971note,Blake.Chwang1974}, the zeroth and first-order terms in $h$ vanish. The Green's function can therefore be expanded to the leading order as
\begin{equation}
	\bm G = h^2 \bm G^{(2)} + \mathcal{O} \left( h^3 \right) . 
\end{equation}
The same expansion also holds for the pressure $P$ and the function $\Psi$. Specifically
\begin{equation}
	P  = h^2 P^{(2)} + \mathcal{O} \left(  h^3 \right) , \qquad \Psi  = h^2 \Psi^{(2)} + \mathcal{O} \left( h^3 \right) ,
\end{equation}
with
\begin{equation}
	\label{eq:p2}
	P^{(2)}=2\int_0^\infty k^2(k+K ) \, e^{-kz} J_0(k\rho) \, \dInt k
\end{equation}
and
\begin{equation}
	\label{eq:psi2}
	\Psi^{(2)}=-2\int_0^\infty k(K+k )  \, e^{-Kz} J_0(k\rho) \, \dInt k \, .
\end{equation}
The first term in each of the integrals can easily be recognized as the second $z$-derivative of the free space solution, Eq.~\eqref{eq:p-psi-infty-fourier}. The solution therefore consists of a force quadrupole and further corrections resulting from the mixed terms mentioned above. We decompose the solutions as
\begin{equation}
	\label{eq:p-psi-decomp}
	P^{(2)}=P^{(2)}_1 +P^{(2)}_2 \, , \qquad    \Psi^{(2)}=\Psi^{(2)}_1+ \Psi^{(2)}_2 \, ,
\end{equation}
where
\begin{equation}
	P^{(2)}_1 = \partial_z^2 P^\infty|_{h=0} \, ,  \qquad \Psi^{(2)}_1= \partial_z^2 \Psi^\infty|_{h=0} \,. 
\end{equation}
In the following, all expressions are evaluated at $h=0$ unless $h$ appears explicitly. 

\subsection{Pressure field}

The first contribution to the pressure in Eq.~\eqref{eq:p-psi-decomp} is obtained as the second derivative of the pressure from Eq.~\eqref{eq:p-psi-inf} and reads
\begin{equation}
	\label{eq:p3}
	P^{(2)}_1 =  6 \, \frac{5z^3-3r^2z}{r^7}\;.
\end{equation}
It corresponds to an axisymmetric source octupole, which can be expressed in terms of a Legendre polynomial as
\begin{equation}
	P_1^{(2)} = \frac{12}{ r^4} \, P_3(z/r) \, .
\end{equation}

For the second contribution, we are not aware of a closed-form analytical solution. By substituting $k = \alpha t$, the second term from the integral in Eq.~\eqref{eq:p2} can be written as
\begin{equation}
	P^{(2)}_2 = 2\alpha^4 \int_0^\infty   t^2 \sqrt{t^2+1} \, e^{-\alpha z t} \,
	J_0(\alpha \rho t) \, \dInt t \, .
	\label{eq:P22}
\end{equation}

In the far-field limit, we have $\alpha z \gg 1$. Therefore, only small values of the integration variable $t$ contribute to the integral. We can expand the expression $\sqrt{t^2+1}$ in a Taylor series 
\begin{equation}
	\label{eq:taylor}
	\sqrt{t^2+1}=\sum_{n=0}^{\infty} \varphi_n t^{2n}
\end{equation}
where
\begin{equation}
	\varphi_n = \frac{(-1)^{n+1} (2n-3)!!}{ 2^{n} \, n! } \, ,
	\label{eq:varphi_n}
\end{equation} 
with $n!!$ denoting a double factorial. The leading coefficients have the values $\varphi_0 = 1$, $\varphi_1 = 1/2$,  $\varphi_2 = -1/8$, $\varphi_3 = 1/16$, $\varphi_4 = -5/128$, etc. 

By inserting Eq.~\eqref{eq:taylor} into Eq.~\eqref{eq:P22}, each term corresponds to a derivative of the free-space pressure Green’s function $P^\infty$ with respect to $z$. This yields the series representation
\begin{equation}
	P^{(2)}_2=-\sum_{n=0}^{{N}} \frac {\varphi_n} {\alpha^{2n-1}} \frac{\partial^{2n+1}}{\partial z^{2n+1}} \, P^\infty \, . 
	\label{eq:P2_2}
\end{equation}
Here, $N$ is the maximum number of terms in the series required for the best approximation. This upper limit is necessary because after the initial convergence, the general term diverges for large $n$. 
Equation~\eqref{eq:P2_2} can be expressed in terms of Legendre polynomials as
\begin{equation}
	P_2^{(2)} = 2\sum_{n=0}^N (2n+2)! \, \frac{ \varphi_n}{\alpha^{2n-1}} \, \frac{P_{2n+2} (z/r)}{r^{2n+3}} \, . 
\end{equation}
For $N=0$, the leading-order term is given by
\begin{equation}
	P_2^{(2)} \simeq 
	\frac{2 \alpha}{r^3} \left( 3 \, \frac {z^2}{r^2}-1\right)
	\label{eq:P2_far-field}
\end{equation}
and has the properties of an axisymmetric source quadrupole. The final asymptotic expression for the far-field pressure is
\begin{equation}
	\label{eq:pfinal}
	P=h^2(P^{(2)}_1+P^{(2)}_2)
	\simeq h^2 \left( 2\alpha \, \frac{3z^2-r^2}{r^5}+ 6 \, \frac{5z^3-3r^2z}{r^7} \right) . 
\end{equation}

In the near-field, $\alpha r \ll 1$, we evaluate Eq.~\eqref{eq:P22} by expanding $\sqrt{t^2+1}$ for large~$t$ as
\begin{equation}
	\sqrt{t^2+1} = t + \frac{1}{2t} - \frac{1}{8t^3} + \mathcal{O} \left( \frac{1}{t^5} \right). 
\end{equation}
We obtain
\begin{equation}
	P_2^{(2)} \simeq 6 \, \frac{5z^3-3r^2z}{r^7} +\alpha^2\, \frac{z}{r^3}   + \frac{\alpha^4}{4} \, 
	\log \left( \alpha (r+z)\right) 
	\qquad (\alpha r \ll 1) \, .
	\label{eq:P2_near-field}
\end{equation}

The first term is identical to $P_1^{(2)}$ from Eq.~\eqref{eq:p3}. Together, they give the pressure $P=12 h^2 (5z^3-3r^2z)r^{-4}$, in agreement with the pressure from Blake's solution \cite{blake1971note} in the limit $h\to 0$. 
The second term represents the leading order correction in a Brinkman fluid. Note that the 
improper integral in Eq.~\eqref{eq:P22} diverges in the third term, but it can still be evaluated up to a constant by first computing $\boldsymbol \nabla P_2^{(2)}$ and then integrating it in space.

\begin{figure}
	\centering
	\includegraphics[width=1\linewidth]{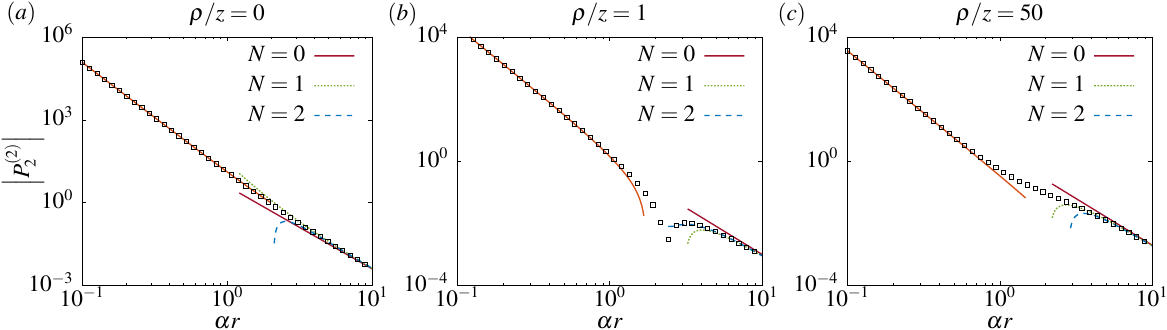}
	\caption{Comparison of the numerically evaluated $P^{(2)}_2$ (symbols) with the asymptotic approximations (solid lines) from Eq.\eqref{eq:P2_far-field} in the far-field limit $\alpha r \gg 1$ and Eq.\eqref{eq:P2_near-field} (up to the order $\mathcal O(\alpha^2)$) in the near-field limit $\alpha r \ll 1$.}
	\label{fig:comparison}
\end{figure}

A comparison between the numerically evaluated $P^{(2)}_2$ and both asymptotic approximations is shown in Fig.~\ref{fig:comparison}.
The far-field prediction from Eq.~\eqref{eq:P2_far-field}, shown in Fig.~\ref{fig:comparison}~(a), agrees well with the numerical integration for $\alpha r \gg 1$. Overall, the best agreement is achieved with $N=1$. Adding additional terms improves the accuracy at $\alpha r \gg 1$, but worsens the divergence in the intermediate range ($\alpha r \lesssim 2$).  The near-field behaviour predicted by Eq.~\eqref{eq:P2_near-field}, illustrated in Fig.~\ref{fig:comparison}~(b), gives a good agreement for $\alpha r \lesssim 1$. Because $P^{(2)}_2$ is just one of the contributions and the others will be given by closed-form analytical expressions, the overall error of any of the approximations in the final result will be less pronounced.

\subsection{Shear-driven flow}

The first contribution to the $\Psi^{(2)}$ in Eq.~\eqref{eq:p-psi-decomp} is obtained as the second derivative of $\Psi^\infty$ from Eq.~\eqref{eq:p-psi-inf} and reads
\begin{equation}
	\label{eq:psi3}
	\Psi^{(2)}_1=  2\left( -(1+\alpha r) +  \frac {z^2}{r^2}(3 + 3\alpha r +(\alpha r)^2) \right) \frac {e^{-\alpha r}}{r^3}\,.
\end{equation}
The second contribution, expressed with a dimensionless integrand, is
\begin{equation}
	\label{eq:psi22}
	\Psi^{(2)}_2 = -2\alpha^3 \int_0^\infty   t^2 \, e^{-\alpha z \sqrt{t^2+1}} \,
	J_0(\alpha \rho t) \, \dInt t \, ,
\end{equation}
which can be evaluated exactly. To do so, we introduce the following convergent improper integral
\begin{equation}
	S_0 = \int_0^\infty
	\frac{1}{\sqrt{t^2+1}} \, e^{-a \sqrt{t^2+1}} J_0(b t) \, \dInt t \, , 
	\label{eq:Sn}
\end{equation}
which can be found in Gradshteyn and Ryzhik~\cite{gradshteyn2014table}, Eq.~6.645.1,  by making the substitution $x = \sqrt{t^2 + 1}$.
Defining $\lambda_\pm = \left( \sqrt{a^2+b^2} \pm a \right) / 2$, we obtain
\begin{equation}
	S_0 = I_0\left( \lambda_- \right)
	K_0 \left( \lambda_+ \right) \, ,
	\label{eq:S0_S1}
\end{equation}
with $K_0$ and $I_0$ denoting the zeroth-order modified Bessel functions of the second and first kind, respectively.
Using higher-order derivatives of $S_0$ with respect to~$a$, denoted by $S_0^{(m)} = \partial^m S_0 / \partial a^m$, the solution of the integral in Eq.~\eqref{eq:psi22} is given by
\begin{equation}
	\Psi^{(2)}_2 = 2\alpha^3(S_0^{(3)} - S_0^{(1)}) \, ,
\end{equation}
evaluated at $a=\alpha z$, $b=\alpha \rho$ and $\lambda_\pm=\alpha (r\pm z)/2$. 
By simplifying the resulting expressions, we obtain
\begin{equation}
	\Psi_2^{(2)} = 
	I_0(\lambda_-) \left[ 
	A_0 K_0(\lambda_+)
	+ \Gamma_+ K_1(\lambda_+)
	\right] 
	+I_1(\lambda_-) 
	\left[ A_1 K_1(\lambda_+) + \Gamma_- K_0(\lambda_+)
	\right] , 
\end{equation}
where
\begin{equation}
	A_0 = \frac{\alpha^2 z}{r^2} \left( 1-\frac{3z^2}{r^2} \right) \, , \quad
	A_1 = - \frac{3\alpha^2 z\rho^2}{r^4} \, , 
\end{equation}
and
\begin{equation}
	\Gamma_\pm = \frac{\alpha}{ r^2}
	\left( 1 \pm \left(1+\alpha^2\rho^2\right) \frac{z}{r}
	-\frac{3z^2}{r^2}
	\mp \frac{3z^3}{r^3} \right) .
\end{equation}

From the asymptotic form of the modified Bessel functions ($K_0(x)\sim \exp(-x)$ and $I_0(x) \sim \exp(x)$) it is clear that $\Psi^{(2)}_2$ decays $\sim \exp(-\alpha z)$, and the same then holds for the velocity field. $\Psi^{(2)}_2$ therefore describes the short-range flows induced by the tractions on the no-slip boundary.

\subsection{Velocity field}
With the knowledge of the two scalar fields, the velocity Green's function is determined as
\begin{equation}
	\bm G^{(2)}= \partial_z^2\bm G^\infty +\alpha^{-2}\boldsymbol \nabla (-P^{(2)}_2 + \partial_z \Psi^{(2)}_2) -\hat{\bm e}_z \Psi^{(2)}_2 \, .
	\label{eq:G2}
\end{equation}
Due to the length of the resulting expression, we do not present the full result here. It can, however, be readily obtained using computer algebra systems or symbolic computation tools.

\subsubsection{Far field}

We now assume $\alpha r \gg  1$, retaining only the terms with $e^{-\alpha z}$, but not those with $e^{-\alpha r}$.  The field $\Psi_1^{(2)}$ does not contribute to the far field and $\Psi_2^{(2)}$ has the asymptotic expansion
\begin{equation}
	\Psi_2^{(2)} \simeq \frac{e^{-\alpha z}}{r^3}
	\left(
	2 + \frac{3\alpha z (3 + \alpha z)}{(\alpha r)^2} 
	+ \frac{15 \alpha z (15 + 15\alpha z + 6 (\alpha z)^2 + (\alpha z)^3)}{4(\alpha r)^4}
	\right) .
\end{equation}

The Green's function for the velocity can now be split up as 
\begin{equation}
	\label{eq:gwh}
	\bm{G}^{(2)}
	= 
	\bm{W}
	+ 
	\bm{H}
	\, , 
\end{equation}
where the first term $\bm W$ represents contributions from $\bm{G}^\infty$ and $P_2^{(2)}$ that correspond to a series of source multipoles. Defining 
\begin{equation}
	\bm{g} = - \boldsymbol{\nabla} \frac 1 r
	= \frac{\rho\, \hat{\bm{e}}_\rho + z \, \hat{\bm{e}}_z}{(\rho^2+z^2)^{3/2}}
\end{equation}
as the flow of a source singularity, 
they can be written compactly as
\begin{equation}
	\bm{W} = \frac{2}{\alpha^2} \frac{\partial^3 \bm{g}}{\partial z^3} 
	-2 \sum_{n=0}^{N} \frac{ \varphi_n}{\alpha^{2n+1}} \, \frac{\partial^{2n+2} \bm{g}}{\partial z^{2n+2}}
\end{equation}
with $\varphi_n$ defined in Eq.~\eqref{eq:varphi_n}. 
In terms of Legendre polynomials, the radial component of $\bm{W}$ can be expressed as
\begin{subequations}
	\begin{equation}
		W_\rho = -\frac{12}{\alpha^2 r^5} \, P_4^1(z/r)
		-2 \sum_{n=0}^N (2n+2)! \, \frac{\varphi_n}{\alpha^{2n+1}} \frac{P_{2n+3}^1 (z/r)}{r^{2n+4}} \, ,
	\end{equation}
	with $P_n^1(z/r) = -(\rho/r)\, P_n'(z/r)$ denoting the associated Legendre function of degree $n$ and order 1, where $P_n'(x)=\dd P_n(x)/\dd x$. 
	The axial component is given by
	\begin{equation}
		W_z = \frac{48}{\alpha^2 r^5 } \, P_4(z/r)
		+ 2 \sum_{n=0}^N (2n+3)! \, \frac{\varphi_n}{\alpha^{2n+1}} \frac{P_{2n+3}(z/r)}{r^{2n+4}} \, .
	\end{equation}
\end{subequations}
By retaining only the first three terms of the series ($N=2$) we obtain
\begin{subequations} \label{eq:W}
	\begin{align}
		W_\rho &\simeq 
		\frac{3\rho}{\alpha r^5}
		\left(
		\frac{10z^2}{r^2}-2
		+\frac{10 z}{\alpha r^2}
		\left( \frac{7z^2}{r^2}-3 \right)
		+\frac{15}{(\alpha r)^2}
		\left(  \frac{21z^4}{r^4}-\frac{14z^2}{r^2}+1  \right) \right), \\
		W_z &\simeq 
		\frac{3}{\alpha r^5}
		\left( 
		\frac{10z^3}{r^2}-6z 
		+ \frac{2}{\alpha}
		\left( \frac{35z^4}{r^4}-\frac{30z^2}{r^2} + 3 \right)
		+ \frac{5z}{(\alpha r)^2}
		\left(  \frac{63z^4}{r^4}-\frac{70z^2}{r^2}+15  \right) \right).
	\end{align}
\end{subequations}

The second term $\bm{H}$ in Eq.~\eqref{eq:gwh} contains the far-field contributions stemming from $\Psi_2^{(1)}$, which can be evaluated exactly, with radial and axial components given in the far field by
\begin{subequations} \label{eq:H}
	\begin{align}
		H_\rho &\simeq \frac{e^{-\alpha z}\rho }{\alpha r^5}
		\left( 6+\frac{15}{(\alpha r)^2} \left( (\alpha z)^2 + 3 \alpha z -3\right)
		\right)
		, \\
		H_z &\simeq -\frac{e^{-\alpha z}}{\alpha^2r^5} 
		\left( 18 + \frac{45 \alpha z}{(\alpha r)^2} (\alpha z+5)
		\right)
		.
	\end{align}
\end{subequations}

Finally, the far-field expression for the velocity is $\bm G=h^2(\bm W +\bm H)$, with $\bm W$ given by Eq.~\eqref{eq:W} and $\bm H$ by \eqref{eq:H}.

\begin{figure}
	\centering
	\includegraphics[width=\linewidth]{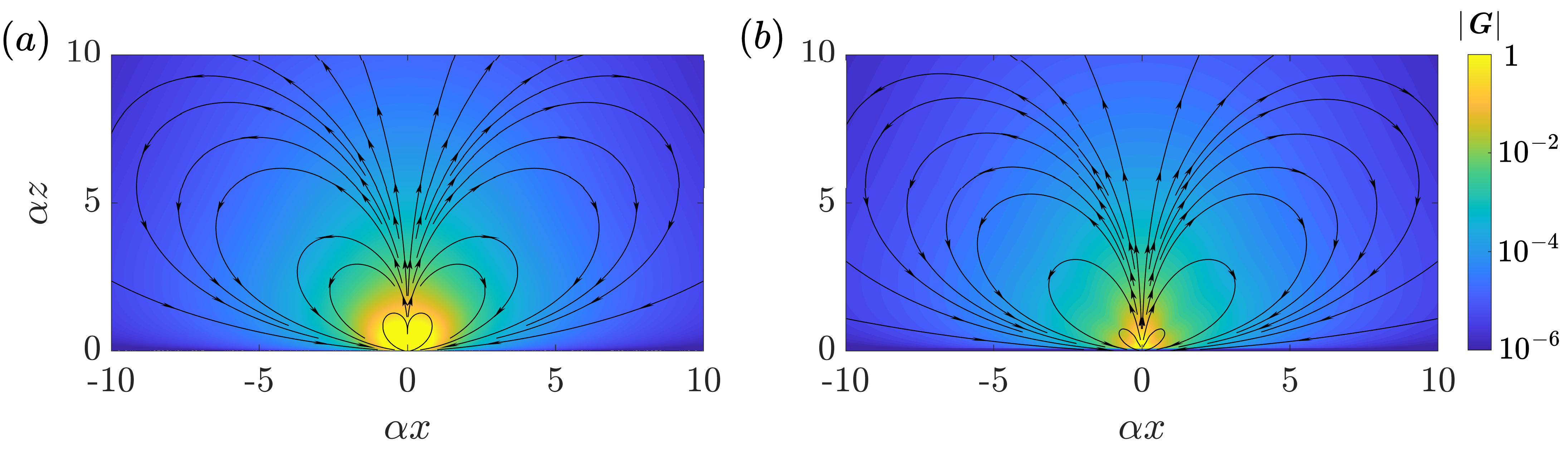}
	\caption{Streamline plot of the axisymmetric flow induced by a Brinkmanlet near a no-slip wall, obtained using (a) the far-field solution from Eqs.~\eqref{eq:W} and \eqref{eq:H} and (b) numerical integration for $\alpha h = 0.1$.
	}
	\label{fig:streamlines}
\end{figure}

A comparison between the far-field flow and the result of numerical integration of Eqs.~\eqref{eq:image-p-psi} and their derivatives is shown in Fig.~\ref{fig:streamlines}, showing near-perfect agreement in the far-field regime.

\subsubsection{Near field}

The velocity can also be analytically evaluated in the near-field limit, where $\alpha r \ll 1$, while still maintaining the condition $r \gg h$.
We obtain
\begin{equation}
	\Psi_2^{(2)} \simeq \frac{2r^2-6z^2}{r^5} +  \alpha^2 \frac{z^2}{r^3} 
	+\frac{\alpha^4}{4} \left( z\left( \frac{1}{4} + \gamma + \log\left( \frac{\alpha (r+z)}{4} \right) \right) - \frac{z^2}{r} \right) .
	\label{eq:Psi2_near}
\end{equation}

Substituting the expressions for $\Psi_2^{(2)}$ from Eq.~\eqref{eq:Psi2_near} and $P_2^{(2)}$ from Eq.~\eqref{eq:P2_near-field}, together with $\bm{G}^\infty$ from Eq.~\eqref{eq:free-space-brinkmanlet}, into Eq.~\eqref{eq:G2}, we obtain
\begin{subequations}
	\begin{align}
		G_\rho^{(2)} &\simeq \frac{6\rho z}{r^7}\left( 3z^2-2\rho^2 \right) + \frac{\alpha^2 \rho z }{2r^5} \left( 2\rho^2-z^2\right) , \\
		G_z^{(2)} &\simeq \frac{6z^2}{r^7} \left( 2z^2-3\rho^2 \right) + \frac{\alpha^2 z^2}{2r^5} \left( \rho^2-2z^2\right) .
	\end{align}
\end{subequations}
The terms of order $\mathcal{O}(\alpha^0)$ coincide with Blake's tensor for a normal force in the limit $r\gg h$ \cite{Blake.Chwang1974}. Terms of the order~$\alpha^2$ then represent the leading correction in the Brinkman flow.

\subsection{Discussion}

A complete solution of the flow problem in the presence of a no-slip wall will also require the Green's function for a force acting parallel to the wall, in which case the height dependence will be $\sim h$. Although the absence of axial symmetry makes the problem more complex, we hope that the solution method developed here can be generalized to parallel forces as well. 

Another open question is whether the same problem can be solved in a 2D Brinkman fluid, which can also be used as an approximation for the Hele-Shaw flow in a thin channel. 

Out results can be used to model hydrodynamic interactions in porous environments where particles or microorganisms move near boundaries. 
A force-free swimmer can, for example, be modeled by placing two oppositely directed point forces asymmetrically relative to the swimmer body~\cite{daddi2018dynamics, menzel16, hosaka2023hydrodynamics}.
In biological contexts near boundaries, such as microorganism motion in gels or tissues, these solutions describe long-range hydrodynamic disturbances.
Moreover, they serve as fundamental building blocks for modeling suspensions near walls or active matter in confined or partially permeable environments.

Although we introduced the Brinkman equation for porous media, our solution remains valid for unsteady flows in the presence of inertia, where we use $\alpha^2= - i \omega /\nu$ with the angular frequency $\omega$ and the kinematic viscosity $\nu=\eta/\rho$~\cite{kim13}.  
For instance, the solution can then be applied for any oscillatory assembly of forces in the proximity of a no-slip boundary. For example, it has been shown that for sufficiently far distances, inertial effects become relevant in the flows driven by a beating cilium \cite{Wei.Tam2019,Wei.Tam2021}. A straightforward extension of our solution could also apply to source singularities near a wall, such as an oscillating gas bubble in the linear regime, or more complex multipole singularities.

\section*{Author contribution}

Both authors jointly designed the research, carried out the analytical calculations, and wrote the manuscript.

\begin{acknowledgments}
A.V.\ acknowledges support from the Slovenian
Research and Innovation Agency (Grants No. P1-0099
and J1-60009).
\end{acknowledgments}

%

\end{document}